\documentclass[runningheads]{llncs}
    
\usepackage{subcaption}
\usepackage{float}
\usepackage{xcolor}
\usepackage{fvextra}
\usepackage{adjustbox}
\usepackage{dblfloatfix}
\usepackage[T1]{fontenc}
\usepackage{graphicx}
\usepackage{amsmath, amssymb}
\usepackage{booktabs}
\usepackage{hyperref}
\usepackage{subcaption, tabularx, cleveref}
\usepackage{placeins}
\usepackage{threeparttable}
\usepackage{tabularx}
\usepackage{booktabs}
    
\usepackage{tikz}
\usetikzlibrary{arrows.meta, positioning, fit, calc, backgrounds}
    
\usepackage[disable]{todonotes} %

\usepackage{enumitem}

\newcommand{\bheading}[1]{{{\textbf{#1.}}}}

\newcommand{\customOrcidID}[1]{{\href{https://orcid.org/#1}{\protect\includegraphics[height=.7em]{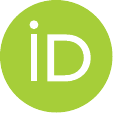}}}}

\begin{document}

\title{Rethinking Neural Decompilation: Candidate Selection as the Hidden Bottleneck}
\titlerunning{Rethinking Neural Decompilation}

\author{Felix Mächtle\,\customOrcidID{0009-0009-2431-0322} \and
Gautham Sriram\,\customOrcidID{0009-0003-9580-8909} \and
Manjil Neupane\,\customOrcidID{0009-0009-3153-368X} \and 
Nils Loose\,\customOrcidID{0009-0003-6243-1623} \and
Thomas Eisenbarth\,\customOrcidID{0000-0003-1116-6973}}

\institute{Institute for IT Security, University of Lübeck, Lübeck, Germany\\
\email{\{f.maechtle, n.loose, thomas.eisenbarth\}@uni-luebeck.de, \\ \{gautham.sriram, manjil.neupane\}@student.uni-luebeck.de}}

\authorrunning{F. Mächtle \textit{et al.}}

\maketitle
    
\begin{abstract}

Reverse engineering obfuscated binaries is a core task in malware analysis and software security auditing. While neural decompilers based on Large Language Models (LLMs) have improved the readability and functional fidelity of recovered code over heuristic tools such as Ghidra and IDA Pro, a substantial gap remains between best-sample performance and practical single-output accuracy, particularly under code obfuscation. We study this gap through the lens of candidate generation and candidate selection. Using the \textit{LLM4Decompile} architecture, we show that ground-truth tokens lie within the model's top-five next-token predictions in 95.6\% of positions on average and that multi-sample decoding reaches a pass@100 of 93\% on HumanEval-Decompile, indicating that correct reconstructions are typically already present in the model's sampling space. We then evaluate a generate-then-select pipeline that ranks 100 candidates per input with text-, syntax-, embedding-, consensus-, and likelihood-based selectors. Negative perplexity emerges as the strongest static selector on HumanEval-Decompile (73.3\% pass@1),  while execution-guided selection nearly closes the oracle gap on ExeBench (86.3\% pass@1). Combined with one epoch of fine-tuning on our newly released ExeBench-Obfuscated dataset containing 362,647 obfuscated OLLVM samples, the same pipeline increases the pass@1 rate for obfuscated binaries to 88.8\%, directly contradicting prior claims that released LLM4Decompile checkpoints are unsuitable for obfuscated inputs.

    \end{abstract}
    
    \section{Introduction}
    
    Binary analysis is a cornerstone of modern software security, enabling critical tasks such as malware detection, vulnerability assessment, and the auditing of proprietary or legacy software~\cite{DBLP:conf/sp/Shoshitaishvili16,DBLP:conf/cav/BrumleyJAS11,DBLP:journals/corr/abs-1905-08325}. A central component of these efforts is \textit{decompilation}, the process of reconstructing high-level source code from machine-executable binaries~\cite{10.1145/3395363.3397370/testing_decompilation_correctness,DBLP:conf/emnlp/TanL0Z24/LLM4Decompile,DBLP:journals/corr/abs-2509-14646}. While traditional decompilers like Ghidra and IDA Pro rely on expert-crafted heuristic rules to recover control flow and program logic, they frequently struggle with the inherent loss of symbolic information during compilation, often producing results that are difficult for human analysts to interpret and impossible to re-execute~\cite{10.1145/3395363.3397370/testing_decompilation_correctness,DBLP:journals/corr/abs-1905-08325}.
    
    The recent success of Large Language Models (LLMs) in the field of program analysis~\cite{DBLP:journals/corr/abs-2601-12951,DBLP:conf/nips/VaswaniSPUJGKP17,DBLP:journals/corr/abs-2504-13676} has also motivated their application to binary decompilation. 
    Models such as \textit{LLM4Decompile}, \textit{SALT4Decompile}, and \textit{ReF Decompile} have demonstrated that treating decompilation as a neural machine translation task can significantly improve code readability and functional correctness \cite{DBLP:conf/emnlp/TanL0Z24/LLM4Decompile,refdecompile,DBLP:journals/corr/abs-2509-14646}. However, despite these advances, a substantial gap remains between apparent code quality and executable correctness. State-of-the-art models frequently generate code that appears  plausible yet fails to satisfy functional test cases. %

This challenge is further exacerbated by code obfuscation techniques such as Control Flow Flattening (FLA) and Bogus Control Flow (BCF), which deliberately distort program structure and obscure semantic relationships. Under these conditions, decompilation models experience severe performance degradation and may even become ineffective at recovering executable high-level code~\cite{DBLP:conf/emnlp/TanL0Z24/LLM4Decompile}.
    
Motivated by this observation, we shift the focus from \textit{whether} neural decompilers can generate correct programs to \textit{whether} they can reliably identify them among multiple plausible candidates. We therefore study neural decompilation as a two-stage problem consisting of candidate generation and candidate selection. To this end, we systematically evaluate a broad range of selection strategies, including text-based, syntax-based, embedding-based, consensus-based, likelihood-based, and execution-guided approaches. Our results demonstrate that candidate selection itself constitutes a major bottleneck in practical neural decompilation: while correct programs are frequently present in the sampled candidate pool, reliably distinguishing them from superficially plausible but functionally incorrect alternatives remains challenging. At the same time, we show that simple selection mechanisms substantially improve single-output accuracy: a likelihood-based selector reaches 73.3\% pass@1 on HumanEval-Decompile~\cite{DBLP:conf/emnlp/TanL0Z24/LLM4Decompile}, improving over the strongest published baseline by 6\%, while execution-guided selection nearly closes the oracle gap on ExeBench at 86.3\% pass@1—all without modifying the underlying decompiler architecture.

Furthermore, we introduce \textit{ExeBench-Obfuscated}, a large-scale dataset of 362{,}647 obfuscated samples generated with common OLLVM transformations. We open-source this dataset on Huggingface and our code on Github: \url{https://github.com/UzL-ITS/rethinking-neural-decompilation}. We demonstrate that multi-candidate selection, combined with targeted fine-tuning on this corpus, substantially improves the robustness of neural decompilers under obfuscation, reaching up to 88.8\% pass@1 on obfuscated binaries—a setting in which released \textit{LLM4Decompile} checkpoints were previously reported as unsuitable~\cite{DBLP:conf/emnlp/TanL0Z24/LLM4Decompile}.

In summary, this work makes the following contributions: (1) a token-level analysis showing that correct reconstructions are largely already present in the model's sampling space; (2) a systematic comparison of selectors that establishes negative perplexity as a new state-of-the-art static selector on HumanEval-Decompile (+6\% over prior work); (3) evidence that generate-then-select, combined with one epoch of fine-tuning, makes off-the-shelf LLM4Decompile usable on obfuscated binaries; and (4) the release of \emph{ExeBench-Obfuscated}, a 362{,}647-sample obfuscation dataset.

\section{Background}
    
    \bheading{Problem Formalization}
    Decompilation is formally defined as the process of mapping a binary executable $x$ back to a high-level source representation $s'$. Given an original source code $s$, the compilation process can be denoted as $x~=~\mathcal{C}(s)$, where $\mathcal{C}$ represents the compiler. A decompiler $\mathcal{D}$ attempts to reconstruct the logic such that $\mathcal{D}(x) = s'$, where $s'$ is expected to maintain high semantic fidelity and functional equivalence to $s$.

\bheading{LLM-based Decompilation}
The current state-of-the-art in neural decompilation, exemplified by
\textit{LLM4De\-compile}, treats decompilation as a sequence-to-sequence
translation problem~\cite{DBLP:conf/emnlp/TanL0Z24/LLM4Decompile}. In contrast
to traditional decompilers such as \textit{Ghidra} or \textit{IDA Pro}, which
recover control-flow and types through hand-crafted
heuristics~\cite{10.1145/3395363.3397370/testing_decompilation_correctness},
a neural decompiler with parameters $\theta$ learns a conditional distribution
over source token sequences $S = (s_1, \ldots, s_T)$ given a binary-derived
input $x$ (e.g., assembly or Ghidra pseudo-code), factorized autoregressively
as
\begin{equation}
    P_\theta(S \mid x) = \prod_{t=1}^{T} P_\theta(s_t \mid x, s_{<t}),
\end{equation}
and is trained by minimizing the negative log-likelihood of $S$ over a
parallel corpus.

At inference, each token $s_t$ is drawn from $P_\theta(s_t \mid x, s_{<t})$
until an end-of-sequence token is produced. Sampling diversity is controlled
by a temperature $\tau \geq 0$ that rescales this distribution: $\tau \to 0$
recovers greedy decoding, while larger $\tau$ flattens it and yields more
diverse outputs.
    
    \bheading{Code Obfuscation}\label{sec:background:code-obfuscation}
    Code obfuscation refers to semantics-preserving transformations intended to make a program difficult to analyze for both humans and automated tools \cite{Obfuscating_history}. In this work, we focus on common techniques provided by the \textit{Obfuscator-LLVM} (OLLVM) framework. Specifically, we use the following techniques:
    \begin{itemize}

       \item \textbf{Control Flow Flattening (FLA):} replaces the original hierarchy with a dispatcher loop that hides program logic.
       \item \textbf{Bogus Control Flow (BCF):} inserts opaque predicates and unreachable code blocks.
       \item \textbf{Instruction Substitution (SUB):} rewrites instructions into longer, semantically equivalent sequences.
       \item \textbf{FLA+Split:} combines flattening with basic-block fragmentation, further obscuring the control-flow graph.
  
  \end{itemize}

    \section{Experimental Methodology}
    \label{sec:methodology}
    
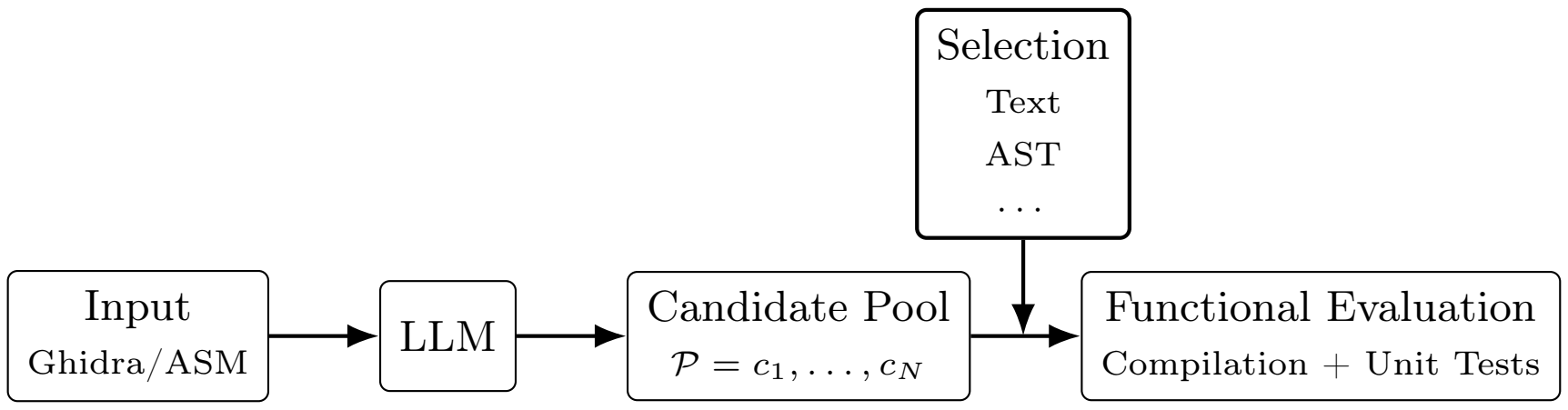
\begin{figure}[t]
    \centering
    \begin{tikzpicture}[
        font=\small,
        >=Latex,
        node distance=2mm and 8mm,
        main/.style={
            draw,
            rounded corners=2pt,
            align=center,
            inner sep=4pt,
            minimum height=8mm
        },
        side/.style={
            draw,
            rounded corners=2pt,
            thick,
            align=left,
            inner sep=4pt
        },
        arrow/.style={->, thick}
    ]
    
    \node[main] (input) {Input\\\scriptsize{Ghidra/ASM}};
    \node[main, right=of input] (llm) {LLM};
    
    \node[main, right=of llm] (pool) {Candidate Pool\\\scriptsize{$\mathcal{P} = c_1, \dots, c_N$}};

    \node[main, right=of pool] (test) {Functional Evaluation\\\scriptsize Compilation + Unit Tests};
    
    \coordinate (midpt) at ($(pool)!0.43!(test)$);
    
    \node[side, align=center, above=7mm of midpt] (scores) {
    Selection\\
    \scriptsize Text\\
    \scriptsize AST\\
    \scriptsize $\cdots$
    };
    
    \draw[arrow] (input) -- (llm);
    \draw[arrow] (llm) -- (pool);

    \draw[arrow] (pool) -- (test);
    \draw[arrow] (scores.south) -| (midpt);
    
    \end{tikzpicture}
    \caption{Overview of our pipeline. The LLM generates a pool of $N$ candidates from Ghidra pseudo-code or assembly, a selector ranks them and the chosen candidate is validated against unit tests.}

    \label{fig:method-pipeline}
\end{figure}

Existing neural decompilation work typically evaluates the final decoded program, thereby mixing the model's ability to assign probability to correct code with the decoding procedure's ability to recover it~\cite{DBLP:conf/emnlp/TanL0Z24/LLM4Decompile,refdecompile,DBLP:journals/corr/abs-2509-14646}. In contrast, we decompose neural decompilation into two stages: a token-level decoder, which turns an input representation into a complete candidate program by autoregressive sampling and a program-level selector, which chooses one output from a pool of independently sampled candidates. %
Figure~\ref{fig:method-pipeline} illustrates the resulting pipeline. Our study is guided by the following three research questions:
    
    \begin{itemize}
        \item RQ1: Is the correct output already present in the model's token-level distribution?
        \item RQ2: Can multi-candidate selection improve single-output decompilation accuracy?
        \item RQ3: Does Multi-Candidate Selection Enable Decompilation of Obfuscated Binaries?
    \end{itemize}

\subsection{Experimental Configurations}
\label{sec:experimental-configurations}

We follow the LLM4Decompile evaluation protocol and reuse its two benchmarks, two input representations, and matching model variants, yielding  the four
configurations C1--C4 listed in Table~\ref{tab:configs}, which we use
throughout the remainder of the paper.

\textbf{Benchmarks.} HumanEval-Decompile~\cite{DBLP:conf/emnlp/TanL0Z24/LLM4Decompile} contains self-contained C functions
translated from HumanEval~\cite{DBLP:journals/corr/abs-2107-03374} and is checked against functional assertions. The
ExeBench dataset~\cite{10.1145/3520312.3534867} contains real-world C
functions extracted from GitHub, including project-specific helpers and
richer control flow, packaged with the I/O wrappers required to compile and
execute each function. All functions in both benchmarks were evaluated at
the four standard GCC optimization levels (O0, O1, O2, O3).

\textbf{Input representations.} We consider  two binary-derived inputs.
\emph{Assembly} listings are produced with \texttt{objdump}, while
\emph{Ghidra pseudo-code} is produced with the Ghidra Headless Analyzer,
which applies Ghidra's decompilation passes to recover a higher-level
representation than raw assembly. For ExeBench, which is distributed as
source code and execution wrappers rather than as a pre-decompiled corpus,
we generate both representations ourselves by compiling each function at
the configured optimization level and then running \texttt{objdump} and
Ghidra Headless on the resulting binary. For the HumanEval-Decompile dataset, we use the samples from~\cite{DBLP:conf/emnlp/TanL0Z24/LLM4Decompile}.

\textbf{Models.} Each input representation is paired with the matching
LLM4De-\\compile variant: \textit{LLM4Decompile-End-6.7B-v1.5} for assembly
inputs and \textit{LLM4-\\Decompile-Ref-9B-v2} for Ghidra pseudo-code inputs.

    \begin{table}[t]
    \centering
    \caption{Experimental configurations following the original \textit{LLM4Decompile} input representations, benchmark family, and model variants.}
    \label{tab:configs}
    \begin{tabularx}{\linewidth}{lXll}
    \toprule
    \textbf{ID} & \textbf{Input} & \textbf{Benchmark} & \textbf{Model} \\
    \midrule
    C1 & Ghidra pseudo-code & HumanEval-Decompile$\hspace{1em}$ & LLM4Decompile-Ref-9B-v2 \\
    C2 & Assembly & HumanEval-Decompile & LLM4Decompile-End-6.7B-v1.5 \\
    C3 & Ghidra pseudo-code & ExeBench & LLM4Decompile-Ref-9B-v2 \\
    C4 & Assembly & ExeBench & LLM4Decompile-End-6.7B-v1.5 \\
    \bottomrule
    \end{tabularx}
    \end{table}

    \subsection{Token-Level Recoverability (RQ1)}
    \label{sec:rq1-method}
    
    For RQ1 we measure whether the ground-truth target tokens are accessible within the model's distribution under a forced prefix, i.e., teacher forcing. At each target position $t$, the model is conditioned on the correct prefix $y_{<t}$ and produces a distribution over the vocabulary $V$, in which we record the rank of the correct token,
    \begin{equation}
        \operatorname{rank}(y_t) = 1 + \left|\left\{v \in V : P_\theta(v \mid x, y_{<t}) > P_\theta(y_t \mid x, y_{<t})\right\}\right|.
    \end{equation}
    A rank of $k$ means the correct token is  contained in the model's top-$k$ predictions. %
    
    To localize prediction difficulty, we additionally group ground-truth tokens into five syntactic categories: identifiers, keywords, literals, operators, and structural tokens. Identifiers include variable, function, type, and field names; keywords include control-flow constructs such as \texttt{if}, \texttt{while}, and \texttt{return}, as well as qualifiers such as \texttt{const} and \texttt{static}; literals include numeric, character, and string constants; operators include arithmetic, logical, comparison, and assignment operators; and structural tokens include braces, parentheses, brackets, semicolons, and commas. For each category, we perform the same rank-based analysis and report the corresponding rank distribution. This allows us to identify the token classes that are the most difficult for the model to recover.
    
\subsection{Candidate Selection (RQ2)}
\label{sec:rq2-method}

Whereas RQ1 examines whether the ground-truth token lies in the model's
distribution at each position, a reverse engineer ultimately requires a
single complete program, not a sequence of token-level predictions.
RQ2 therefore lifts the analysis from tokens to programs and asks whether a functionally correct candidate can be identified without access to the
ground truth. %

\bheading{Candidate generation}
\label{sec:rq2-generation}
For each input $x$ we draw multiple  candidates with varying temperatures $\tau \in \{0.0, 0.3, 0.5, 0.7, 0.9, 1.0\}$, mixing
high-confidence completions with progressively more diverse
samples.
The candidate pool $\mathcal{P}$ is fixed before any selector runs, so any
difference between selectors is attributable to the selection criterion
rather than to sampling variance in the generator.

\bheading{Selection signals}
\label{sec:rq2-selection}
We evaluate selectors from four families. Each scores against a compilable candidate $c \in \mathcal{P}$: similarity to
the input $x$, agreement within the pool $\mathcal{P}$, the
decompiler's own likelihood of $c$, or runtime behavior of the compiled candidate. 
Two baselines complete the comparison: \emph{Random} selection over $\mathcal{P}$ provides a no-information lower
reference, while an \emph{oracle} that returns any candidate passing
all functional tests serves as an upper bound over the fixed pool and
is therefore not deployable. All selectors are listed and 
described in Appendix~\ref{app:rq2-selectors}.

\subsection{Obfuscation Dataset}
\label{sec:main:obfuscation-dataset}

To evaluate whether candidate selection remains effective beyond standard, unobfuscated binaries, we require a paired dataset that exposes neural decompilers to semantics-preserving obfuscation while retaining ground-truth source code.

To the 
best of our knowledge, no such dataset exists at the scale required 
for fine-tuning LLM-based decompilers. %
We therefore construct an obfuscation dataset based on functions from ExeBench, named \textit{ExeBench-Obfuscated}. 
For each sample, the target function and its dependencies are combined into a compilable source file and compiled twice: once unobfuscated (used in C2/C4 of RQ2) and once through an OLLVM toolchain with one of the four obfuscation passes described in \Cref{sec:background:code-obfuscation} (used in RQ3).

Throughout this process, we preserve the original function code, dependency information and compilation outcomes. To improve dataset quality, we remove empty or trivial functions, i.e., functions whose bodies are empty or reduce to a bare \texttt{return;} statement, since such cases provide little meaningful signal for decompilation-based learning.

Finally, the generated obfuscated samples are mapped back to the original ExeBench samples. This restores repository-level metadata and execution-related fields, including the original source path, I/O specification, execution wrapper, and  identifier. Following Loose \emph{et al.}~\cite{loose2026a}, we do not split the training and test partition randomly but by project. This prevents data snooping within a project. \textit{ExeBench-Obfuscated}  contains 287{,}790 training samples and 74{,}857 test samples, leading to a total of 362{,}647 samples.

    \subsection{Evaluation Metrics}
    \label{sec:evaluation-metrics}

Following prior work on neural decompilation~\cite{DBLP:conf/emnlp/TanL0Z24/LLM4Decompile},
our primary metric is \emph{functional correctness}. A candidate $c$ is counted
as correct for an input $x$ if it (i) compiles successfully under the
evaluation harness and (ii) passes all associated functional tests within a
two-second timeout; candidates that fail to compile, time out, or fail any
assertion are counted as incorrect.

\section{Evaluation}
    
We present results for each research question across the four configurations described in Section~3.1 (Table~\ref{tab:configs}).

\subsection{RQ1: Is the Correct Token in the Sampling Space?}

RQ1 examines whether decompilation failures arise because the model cannot represent the correct output, or because decoding fails to recover it. To this end, we analyze the model's next-token distribution under teacher forcing and measure how often each ground-truth token appears among the model's highest-probability predictions.

    \bheading{Setup}
    We applied the teacher-forced rank analysis from Section~\ref{sec:rq1-method} to each of the four configurations using 100 randomly sampled functions per configuration. At every target position, we recorded the rank of the ground-truth token under the model's distribution. The token-category ablation followed the five-category grouping defined in Section~\ref{sec:rq1-method}.

\begin{table}[t]
\centering
\caption{RQ1: Token rank analysis across configurations. For each configuration, we report the percentage of tokens where the correct token appears at rank~1 (exact top prediction), within the top~5, and within the top~10. }
\label{tab:rq1-summary}
\begin{tabular}{llllrrrr}
\toprule
\textbf{Config} & \textbf{Input} & \textbf{Benchmark} & \textbf{Model} & \textbf{Rank\,=\,1} & \textbf{$\leq$\,5} & \textbf{$\leq$\,10} & \textbf{Mean} \\
\midrule
C1 & Ghidra & HumanEval & 9B-v2 & 80.9\% & 93.7\% & 95.6\% & 15.8 \\
C2 & ASM & HumanEval & 6.7B-v1.5 & 87.5\% & 97.2\% & 97.9\% & 5.5 \\
C3 & Ghidra & ExeBench & 9B-v2 & 87.3\% & 97.1\% & 98.1\% & 6.2 \\
C4 & ASM & ExeBench & 6.7B-v1.5 & 82.2\% & 94.5\% & 96.2\% & 10.1 \\
\midrule
\textit{Avg.} & -- & -- & -- & \textit{84.5\%} & \textit{95.6\%} & \textit{97.0\%} & \textit{9.4} \\
\bottomrule
\end{tabular}
\end{table}
    
\bheading{Results}
Table~\ref{tab:rq1-summary} reports the rank distribution per configuration. The ground-truth token is the top prediction (rank~=~1) in 80.9--87.5\% of positions and falls within the top five in 93.7--97.2\% of positions. Averaged across the four configurations, the ground-truth token is at rank~=~1 in 84.5\% of positions, within the top five in 95.6\%, and within the top ten in 97.0\%. Tokens with rank~>~10 account for only 1.9--4.4\% of positions. The mean rank, however, varies substantially across configurations, from 5.5 on C2 to 15.8 on C1. %
Table~\ref{tab:rq1-ablation} reports the rank~=~1 rates per syntactic category. Identifiers reach rank~=~1 in 70.3--83.9\% of positions, keywords in 71.2--75.5\%, literals in 76.4--92.6\%, operators in 85.9--90.3\%, and structural tokens in 86.1--90.4\%. %

\bheading{Discussion}
Across all four configurations the correct token sits within the model's top five predictions in 93\%--97\% of positions, so the failures of greedy decoding cannot be blamed on a missing target distribution. The residual error concentrates where one would expect: identifiers and the qualifier subset of keywords (\texttt{const}, \texttt{static}, etc.) carry information that compilation discards, while operators and structural tokens, which are pinned down by the recovered control flow and expression structure, are recovered almost perfectly.

Two configuration-level effects qualify this picture. The Ghidra-to-assembly switch degrades identifier recovery on ExeBench but halves the literal gap on HumanEval, which we attribute to the short, context-poor HumanEval functions rather than to assembly input in general. Second, C1's mean rank of 15.8 — well above the 5.5–10.1 elsewhere — shows that, when the Ghidra-refined model makes errors on HumanEval, it does so definitively, with probability mass leaking far down the vocabulary.

\begin{table}[t]
\centering
\caption{RQ1: Token rank analysis by syntactic category. Rank\,=\,1 (\%) for grouped AST node types across configurations. Identifiers and keywords are consistently the hardest categories, while structural tokens and operators are near-perfectly predicted.}
\label{tab:rq1-ablation}

{\setlength{\tabcolsep}{7pt}
\begin{tabular}{lccccc}
\toprule
\textbf{Token Category} & \textbf{C1} & \textbf{C2} & \textbf{C3} & \textbf{C4} & \textbf{Avg} \\
\midrule
Identifiers & 77.9\% & 72.9\% & 83.9\% & 70.3\% & 76.3\% \\
Structural  & 88.8\% & 86.1\% & 90.4\% & 88.5\% & 88.5\% \\
Operators   & 86.6\% & 85.9\% & 90.3\% & 86.7\% & 87.4\% \\
Keywords    & 75.5\% & 73.7\% & 74.8\% & 71.2\% & 73.8\% \\
Literals    & 91.5\% & 76.4\% & 92.6\% & 90.7\% & 87.8\% \\
\bottomrule
\end{tabular}}
\end{table}

    \subsection{RQ2: Can Selection Improve Pass@1?}

\begin{figure}[t]
    \centering
\includegraphics[width=\linewidth]{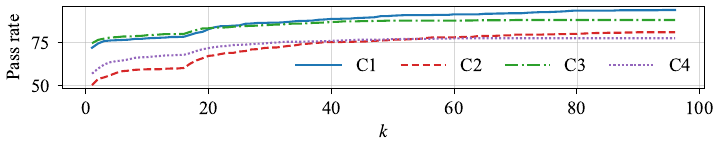}
    \caption{Oracle pass rate versus candidate pool size $k$. Curves saturate
around $k \in [60, 80]$, motivating our choice of $N = 100$.}
    \label{fig:oracle-pass-vs-pool-size}
\end{figure}
    
RQ1 shows that the model assigns a high probability to the ground-truth tokens. However, since some information is lost during compilation, it must be estimated. In the autoregressive manner in which these models work, this means that mistakes are made, if lost information  is guessed wrong~\cite{DBLP:conf/nips/VaswaniSPUJGKP17}. RQ2 lifts the analysis from individual tokens to complete programs by asking whether a selection function applied to a pool of sampled candidates can reliably identify a functionally correct one. 

\bheading{Pre-study: candidate pool size}
Before comparing selectors, we first fix a candidate pool size $N$ large
enough that the oracle pass rate has effectively saturated. %
To pick $N$ in a
principled way, we sample $100$ candidates per input under the
temperature schedule $\tau \in \{0.0, 0.3, 0.5, 0.7, 0.9, 1.0\}$ and
compute the oracle pass rate for every pool size $k \in \{1, \dots, 100\}$.
Figure~\ref{fig:oracle-pass-vs-pool-size} shows the resulting curves. All four
configurations rise steeply over the first $\sim$20 candidates and flatten
between $k\!=\!60$ and $k\!=\!80$; beyond that point, additional samples
yield at most marginal gains. However, the saturation point itself varies
across configurations. We therefore fix $N = 100$ for the remainder of
this section. This choice is intentionally conservative: it ensures that,
across all four configurations, the candidate pool is saturated with
respect to the model's sampling distribution. %
    
    \bheading{Setup}
    We instantiate the generate-then-select pipeline of \Cref{sec:rq2-generation} on all four configurations from ~\Cref{sec:experimental-configurations}. For HumanEval-Decompile (C1, C2) we use all 656 inputs (164 functions $\times$ 4 optimization levels); for ExeBench (C3, C4) we sample 1000 inputs uniformly. For each input we draw $N=100$ candidate programs and score every candidate with each selector from Section~\ref{sec:rq2-selection}. However, not all samples are compilable. Therefore, the selection analysis excludes inputs for which no compilable candidate could be generated, leaving 634 (C1), 640 (C2), 774 (C3), and 428 (C4) effective inputs. 

\bheading{Results}
\Cref{tab:rq2-selection} reports pass@1 for every selector across the four configurations. The oracle ranges from 75.7\% (C4) to 92.7\% (C1), and the gap between random selection and the oracle, the empirical size of the selection bottleneck, ranges from 16.3 (C4) to 33.4 (C2) percentage points.

Among static selectors, negative perplexity is the strongest on three of four configurations: C1 (73.3\%, +5.1\,pp over random), C2 (55.0\%, +8.0\,pp), and C4 (65.1\%, +5.7\,pp). On C3, the result is different:  perplexity reaches  only 70.4\% against a random baseline of 69.5\%, a gap that lies within sampling noise, while MBR consensus took the lead at 72.5\% (+3.0\,pp) and the two compile-filtered majority-vote variants followed closely at 72.2\%. MBR consensus is  otherwise consistently behind perplexity on the HumanEval configurations, closing 14\% of the oracle--random gap on C1 (71.6\%) and 15\% on C2 (51.9\%). Similarity-based selectors are  uniformly weaker than perplexity on HumanEval; on C3 the closest similarity selectors (Edit Similarity 69.4\%, Trigram 69.3\%, Bigram 69.0\%) cluster around the random baseline without clearly beating it. The two majority-vote variants that include non-compiling candidates (rows c, d) collapse on ExeBench, falling to 37.4\% on C3 and 14.9--15.0\% on C4, well below random.

Execution-guided selection, the only dynamic selector  evaluated, shows a sharply different profile. On HumanEval it is  middling (70.8\% on C1, 51.8\% on C2), trailing perplexity, MBR, and both compile-filtered majority votes. On ExeBench, by contrast, it is the strongest selector by a wide margin, reaching 86.3\% on C3 and 75.4\% on C4---in both cases within 0.5\,pp of the oracle.

Candidate generation costs 0.126\,s per sample on average, so a 100-candidate pool adds roughly 12.6\,s of inference time per input on top of the selector cost.

\begin{table}[t]
    \centering
    \caption{RQ2: Average pass@1 by selector across five seeds on a pool of $N=100$ candidates per input. \textit{Oracle} is the upper bound that picks any passing candidate. Avg. is the unweighted mean over C1–C4; per-seed standard deviations are reported per configuration.}
    \label{tab:rq2-selection}
    
    {\begin{threeparttable}
    \setlength{\tabcolsep}{5pt}
    \scriptsize
    \begin{tabularx}{1\linewidth}{Xllllr}
    \toprule
    \textbf{Selector}              & \textbf{C1}   & \textbf{C2}   & \textbf{C3}   & \textbf{C4} & \textbf{Avg.} \\
    \midrule
\textit{Oracle} & $\mathit{92.7 \pm 1.0}$ & $\mathit{80.4 \pm 0.6}$ & $\mathit{86.8 \pm 0.6}$ & $\mathit{75.7 \pm 1.6}$ & $\mathit{83.9}$ \\
Execution-Guided & $70.8 \pm 0.4$ & $51.8 \pm 0.9$ & $\mathbf{86.3 \pm 0.8}$ & $\mathbf{75.4 \pm 1.6}$ & $\mathbf{71.1}$ \\
Neg. Perplexity & $\mathbf{73.3 \pm 0.5}$ & $55.0 \pm 1.2$ & $70.4 \pm 2.0$ & $65.1 \pm 2.8$ & $66.0$ \\
MBR Consensus & $71.6 \pm 0.5$ & $\mathbf{51.9 \pm 0.7}$ & $72.5 \pm 1.2$ & $62.3 \pm 2.2$ & $64.6$ \\
Majority Vote\tnote{a} & $71.2 \pm 0.4$ & $51.5 \pm 0.6$ & $72.2 \pm 1.4$ & $61.5 \pm 2.5$ & $64.1$ \\
Majority Vote\tnote{b} & $71.0 \pm 0.5$ & $51.5 \pm 1.0$ & $72.2 \pm 1.7$ & $61.6 \pm 2.6$ & $64.1$ \\
Trigram Similarity & $66.2 \pm 1.3$ & $51.6 \pm 0.9$ & $69.3 \pm 0.5$ & $60.9 \pm 3.7$ & $62.0$ \\
Random & $68.2 \pm 1.2$ & $47.0 \pm 1.2$ & $69.5 \pm 1.4$ & $59.4 \pm 1.9$ & $61.0$ \\
UniXcoder (Fine-tuned) & $66.8 \pm 1.2$ & $46.4 \pm 0.7$ & $68.0 \pm 1.2$ & $59.5 \pm 3.2$ & $60.2$ \\
Bigram Similarity & $65.7 \pm 1.3$ & $44.9 \pm 1.1$ & $69.0 \pm 0.9$ & $60.6 \pm 3.4$ & $60.0$ \\
RepE Correctness & $66.0 \pm 2.3$ & $45.3 \pm 1.6$ & $68.6 \pm 1.1$ & $58.8 \pm 4.2$ & $59.7$ \\
Ensemble (Text) & $66.5 \pm 1.6$ & $42.2 \pm 0.7$ & $68.9 \pm 0.3$ & $60.8 \pm 3.5$ & $59.6$ \\
Edit Similarity & $64.8 \pm 1.9$ & $43.0 \pm 2.5$ & $69.4 \pm 1.2$ & $61.4 \pm 2.6$ & $59.6$ \\
Identifier Similarity & $60.2 \pm 0.9$ & $46.8 \pm 1.4$ & $68.5 \pm 1.6$ & $61.6 \pm 4.2$ & $59.3$ \\
Token F1 & $64.2 \pm 0.8$ & $42.6 \pm 1.2$ & $67.9 \pm 1.5$ & $60.3 \pm 3.1$ & $58.8$ \\
AST Node Similarity & $61.0 \pm 1.2$ & $39.9 \pm 1.5$ & $67.9 \pm 0.6$ & $58.7 \pm 3.3$ & $56.9$ \\
UniXcoder (Base) & $62.1 \pm 1.6$ & $38.5 \pm 0.9$ & $67.5 \pm 1.5$ & $59.3 \pm 3.5$ & $56.9$ \\
Control Flow Similarity & $58.4 \pm 2.3$ & $35.9 \pm 0.7$ & $67.9 \pm 1.4$ & $61.5 \pm 2.5$ & $55.9$ \\
Majority Vote\tnote{c} & $65.2 \pm 0.5$ & $48.0 \pm 0.3$ & $37.4 \pm 0.6$ & $15.0 \pm 0.2$ & $41.4$ \\
Majority Vote\tnote{d} & $65.0 \pm 0.4$ & $47.8 \pm 0.5$ & $37.4 \pm 0.8$ & $14.9 \pm 0.2$ & $41.3$ \\
    \midrule
    \multicolumn{6}{l}{\emph{Related Work (Scores taken from publication)}} \\
    SK$^2$ Decompile~\cite{tan2026skdecompile} & 69.0 & -- & -- & -- & 69.0\\ 
    CodeInverter~\cite{liu2025codeinverter} & -- & 62.1 & -- & -- & 62.1 \\
    ReF Decompile~\cite{refdecompile} & -- & 61.4 & -- & -- & 61.4\\
    HELIOS~\cite{DBLP:journals/corr/abs-2601-14598} & 55.9 & -- & -- & -- & 55.9\\
    SccDec +  FAE~\cite{DBLP:conf/emnlp/FengT0M00ZC24} & -- & 52.4 & -- & -- & 52.4\\
    SALT4Decompile~\cite{DBLP:journals/corr/abs-2509-14646} & -- & 58.7 & -- & 26.4 & 42.5\\
    \bottomrule
    \end{tabularx}\scriptsize
    a) Normalized, Compile Only, b) Exact, Compile Only, c) Normalized, All, d) Exact, All.
    \end{threeparttable}}
\end{table}

\bheading{Discussion}
The selector ranking is benchmark-dependent in a way that maps onto the candidate distribution each benchmark induces. On HumanEval-Decompile  perplexity is the strongest static signal, with consensus-based selectors a step behind. On ExeBench, project-specific helpers and richer control flow push the model off its training distribution, flattening likelihoods until they no longer discriminate. This likely explains why perplexity ties with random selection on C3, and why consensus-based selectors such as MBR and compile-filtered majority vote take over there. %

Execution-guided selection is not affected by likelihood collapse because it evaluates candidates based on actual executions of the harness, and on ExeBench it nearly closes the oracle gap. On HumanEval-Decompile it underperforms perplexity and MBR. We attribute this to the lower complexity of HumanEval-Decompile, which weakens the discriminative power of the output-guided signal. Execution-guided selection is therefore most useful where precise input--output pairs are available. However, this condition is rarely met in real reverse-engineering settings, where static selectors offer a more realistic alternative.

Orthogonally, Ghidra-input configurations (C1, C3) lead their assembly counterparts, indicating that the Ghidra front-end's symbol- and structure-level recovery still adds value when a strong selector follows. The added inference cost is modest: at 0.126\,s per sample, generating a 100-candidate pool takes roughly 12.6\,s per input, which is negligible compared to the manual effort a reverse engineer would otherwise spend on the same function.

\subsection{RQ3: Decompilation of Obfuscated Binaries}
\label{sec:rq3-results}

RQ1 and RQ2 focus on standard decompilation benchmarks, but real-world reverse
engineering frequently targets binaries that have been deliberately obfuscated to
hinder static analysis. RQ3 therefore asks whether the generate-then-select
pipeline of RQ2 enables decompilation in this setting. The question is
particularly relevant because the released \textit{LLM4Decompile} checkpoints
were not trained on obfuscated binaries, and the original work explicitly
states that they are unsuitable for such
inputs~\cite{DBLP:conf/emnlp/TanL0Z24/LLM4Decompile}. We therefore evaluate
the off-the-shelf checkpoint and an
obfuscation-aware fine-tuned variant  with the
generate-then-select pipeline from RQ2.

\bheading{Setup}
We use \textit{LLM4Decompile-Ref-9B-v2} as the base for fine-tuning because it achieved
the strongest pass@1 in configuration~C1 of RQ2 and operates on Ghidra
pseudo-code, the stronger input representation in our evaluation. For fine-tuning we train for a single epoch on \textit{ExeBench-Obfuscated}  (see~\Cref{sec:main:obfuscation-dataset}). Evaluation uses 1000 random  obfuscated samples covering four obfuscation classes (bcf, $n=233$; fla, $n=208$; fla+split, $n=251$; sub, $n=308$) and four optimization levels (O0, $n=219$; O1, $n=262$; O2, $n=235$; O3, $n=284$). As selection
strategies we use \emph{Perplexity} and \emph{Execution-guided}, the best static and
dynamic mechanisms from RQ2. %
In addition to pass@1, we report Levenstein similarity and CodeBLEU~\cite{DBLP:journals/corr/abs-2009-10297} to capture how closely each recovered candidate matches the ground-truth source beyond strict functional equivalence.
    
\bheading{Results}
\Cref{tab:rq3-overall} shows that fine-tuning consistently improves
performance under both selection strategies. With perplexity selection, the
original model reaches 58.1\% pass@1, 88.0\% compile rate, 38.4\% Levenshtein,
and 35.9\% CodeBLEU, while the fine-tuned model improves to 77.1\%, 94.7\%,
54.4\%, and 44.4\%, respectively. Execution-guided selection further increases
pass@1 to 79.5\% for the original model and 88.8\% for the fine-tuned model,
with the fine-tuned checkpoint also achieving the best overall similarity
scores.
    
      \begin{table}[t]
      \centering
      \caption{RQ3: Obfuscated decompilation results with perplexity and execution-guided selection over a 100-candidate pool per input. Original is the off-the-shelf \textit{LLM4Decompile-Ref-9B-v2}; Fine-tuned is the same checkpoint after training one epoch on \textit{ExeBench-Obfuscated}.}
      \label{tab:rq3-overall}
      \begin{tabularx}{\linewidth}{Xcccc}
      \toprule
      \textbf{Setting} & \textbf{Pass@1} & \textbf{Compile} & \textbf{Levenshtein} & \textbf{CodeBLEU} \\
      \midrule
      Original + Perplexity       & 58.1\% & 88.0\% & 38.4\% & 35.9\% \\
      Fine-tuned + Perplexity     & 77.1\% & 94.7\% & 54.4\% & 44.4\% \\
      Original + Execution-guided & 79.5\% & 88.0\% & 39.8\% & 36.2\% \\
      Fine-tuned + Execution-guided & 88.8\% & 94.7\% & 55.0\% & 44.6\% \\
      \bottomrule
      \end{tabularx}
      \end{table}
    
Table~\ref{tab:rq3-breakdown} reports the breakdown by obfuscation class
and by optimization level. The fine-tuned model exceeds the original on
every class and every optimization level. The largest absolute pass@1 gain
among classes is at \textit{bcf} (+13.3pp), closely followed by \textit{fla+split} (+13.1pp);
the smallest is at \textit{sub} (+4.5pp). Among optimization levels, the largest
gain is at O3 (+11.6pp) and the smallest at O0 (+6.8pp). The pass@1 spread
of the original model across classes is 5.8 points (77.3\% at \textit{bcf} versus
83.1\% at \textit{sub}) and across optimization levels is 8.8 points (75.7\% at O3
versus 84.5\% at O0). After fine-tuning these spreads are 4.7 points
(86.1--90.8\%) and 5.3 points (86.0--91.3\%) respectively.

    \begin{table}[t]
    \centering
    \caption{RQ3: Pass@1 of the original and fine-tuned model with execution guided selection, broken down by obfuscation class and by optimization level. All values are pass@1 percentages.
    }
    \label{tab:rq3-breakdown}
    \begin{tabular}{lcccc}
    \toprule
    \textbf{Group} & \textbf{Original} & \textbf{Fine-tuned} & \textbf{$\Delta$} & \textbf{$n$} \\
    \midrule
    \multicolumn{5}{l}{\emph{By obfuscation class}} \\
    \quad bcf       & 77.3\% & 90.6\% & +13.3pp & 233 \\
    \quad fla       & 78.8\% & 86.1\% & +7.2pp & 208 \\
    \quad fla+split & 77.7\% & 90.8\% & +13.1pp & 251 \\
    \quad sub       & 83.1\% & 87.7\% & +4.5pp  & 308 \\
    \midrule
    \multicolumn{5}{l}{\emph{By optimization level}} \\
    \quad O0 & 84.5\% & 91.3\% & +6.8pp & 219 \\
    \quad O1 & 81.7\% & 90.8\% & +9.2pp & 262 \\
    \quad O2 & 77.0\% & 86.0\% & +8.9pp & 235 \\
    \quad O3 & 75.7\% & 87.3\% & +11.6pp & 284 \\
    \bottomrule
    \end{tabular}
    \end{table}
    
\bheading{Discussion}
Selection makes the off-the-shelf checkpoint usable on inputs it was never trained on, directly contradicting the disclaimer in the original \textit{LLM4Decompile} paper~\cite{DBLP:conf/emnlp/TanL0Z24/LLM4Decompile}. Execution-guided
selection retains its RQ2 advantage whenever a harness is available, and
fine-tuning compounds with both selectors rather than substituting for
them. The class-level breakdown is consistent with the structure of the
obfuscations: \textit{sub} preserves control-flow shape and therefore yields Ghidra
output closest to a non-obfuscated trace, leaving the least headroom for an
obfuscation-aware corpus to exploit, whereas the three control-flow
transforms benefit substantially more. Across optimization levels, fine-tuning
helps most where the original model struggled most, compressing the O0--O3
spread from 8.8 to 5.3 points. Moreover, because the corpus is split by project (\Cref{sec:main:obfuscation-dataset}),
the reported numbers approximate cross-project generalization rather than
training data  memorization.

\subsection{Case Study: Malware Reverse Engineering}
\label{sec:case-study}

To illustrate the pipeline on real-world adversarial code, we applied it to a
single function from the Mirai IoT botnet, a real-world malware sample~\cite{DBLP:conf/uss/AntonakakisABBB17/Mirai-Botner}.

We focused on \texttt{toggle\_obf}, the routine that XOR-decrypts (and
re-encrypts) entries in Mirai's encrypted string table. We chose this
function because it is the most analyst-relevant single function in the
binary: a correct recovery directly exposes the XOR key, which in turn
unlocks every C2 domain, command string, and kill-list entry referenced
elsewhere in the bot. We compiled the leaked  source code with
\texttt{gcc -O2}, obtained Ghidra pseudo-code
for \texttt{toggle\_obf} via the Headless Analyzer, and ran three pipelines on
the same input: (i) Ghidra alone (the standard analyst baseline), (ii) the
off-the-shelf \textit{LLM4Decompile-Ref-9B-v2} checkpoint with the
generate-then-select configuration from RQ2 ($N=100$ candidates, perplexity
selection), and (iii) the fine-tuned checkpoint from RQ3 under the same
configuration. The outputs are shown side-by-side in the Appendix in ~\Cref{fig:case-study}
alongside the leaked ground-truth source.

The four reconstructions form a clear progression in source-level fidelity. The ground truth expresses \texttt{toggle\_obf} as a byte-wise XOR over \texttt{val->val}, with the 32-bit \texttt{table\_key} split into four single-byte subkeys and an optional \texttt{DEBUG} toggle. Ghidra recovers this behavior but buries it under pointer arithmetic, synthetic types, and inlined struct offsets (CodeBLEU similarity: 11.3). The off-the-shelf LLM4Decompile checkpoint with perplexity selection lifts the loop back into readable C and recovers the four masked XOR steps, though field names and struct layout remain approximate (CodeBLEU similarity: 20.1). The fine-tuned variant from \Cref{sec:rq3-results}, with the same  selection, tracks the ground truth most closely (CodeBLEU similarity: 27.0): it reproduces the four single-byte key extractions through explicit shift-and-mask expressions matching the original, and exposes the target buffer through named pointer variables rather than offset-indexed casts. Residual differences are confined to variable types and the exact form of struct field access.

\section{Related Work}
    
\bheading{Traditional and Neural Decompilation}
Binary decompilation is the process of reversing machine code into high-level source code %
\cite{DBLP:conf/sp/Shoshitaishvili16,DBLP:journals/corr/abs-1905-08325}. Traditional decompilers, such as \textit{Ghidra} and \textit{IDA Pro}, rely on hand-crafted heuristic rules and control-flow analysis to reconstruct source structures. Thus, these tools often produce output that is difficult to read, lacks meaningful variable names, and is frequently non-recompilable due to the inherent loss of symbolic information during compilation \cite{10.1145/3395363.3397370/testing_decompilation_correctness}.
    
With the rise of Large Language Models (LLMs), decompilation has been reformulated as a neural machine translation task~\cite{DBLP:journals/corr/abs-1905-08325}. Early neural approaches utilized sequential models, e.g., RNNs, to learn mappings between assembly and source code \cite{DBLP:conf/wcre/KatzRS18}. Recently, \textit{LLM4Decompile} established a new baseline by releasing a series of open-source models (1.3B to 22B) fine-tuned on billions of tokens \cite{DBLP:conf/emnlp/TanL0Z24/LLM4Decompile}. %
Several recent works have attempted to improve upon the \textit{LLM4Decompile} foundation. \textit{SccDec} introduced fine-grained alignment enhancement (FAE) to better preserve original functionality \cite{DBLP:conf/emnlp/FengT0M00ZC24}. \textit{ReF Decompile} proposed a relabeling strategy for jump addresses and a function call mechanism to retrieve missing variable information from binaries \cite{refdecompile}. Most recently, \textit{SALT4Decompile} introduced the Source-level Abstract Logic Tree (SALT) to guide LLMs through complex control flows~\cite{DBLP:journals/corr/abs-2509-14646}. 

However, all of these efforts target the \emph{generation} stage. %
Our work is orthogonal: we leave the underlying decompiler unchanged and instead address \emph{selection} from a pool of sampled candidates. The two directions compose by construction, i.e., any improved generator slots directly into our generate-then-select pipeline. 
    
\bheading{Obfuscation and Robustness Challenges}
Code obfuscation has long been a fundamental obstacle for binary 
analysis~\cite{DBLP:journals/csur/SchrittwieserKK16}. For neural decompilers, the impact is severe: the 
original \textit{LLM4Decompile} reports a re-executability drop of more 
than 70\% under FLA and BCF and concludes that its checkpoints are 
unsuitable for obfuscated inputs~\cite{DBLP:conf/emnlp/TanL0Z24/LLM4Decompile}. 
Tkachenko \textit{et al.} evaluate seven commercial LLMs against BCF, IS, FLA, 
and their combination, finding combined obfuscation defeats every model 
tested~\cite{DBLP:journals/corr/abs-2505-19887}. The problem is compounded 
by a lack of training data: large source-aligned corpora such as \textit{ExeBench}~\cite{10.1145/3520312.3534867}, \emph{Conan}~\cite{DBLP:conf/acns/MachtleSLSE25/OCEAN} or  \textit{Decompile-Bench}~\cite{tan2025decompilebenchmillionscalebinarysourcefunction} contain no obfuscated samples. \textit{BinKit}~\cite{DBLP:journals/tse/KimKCSK23} and the Quarkslab diffing dataset~\cite{quarkslab:diffing_obfuscation_dataset:2025} do provide obfuscated binaries at scale, but  do not align obfuscated binaries with their original source functions, required for decompiler training. 
Notably, concurrently with our work, a second dataset derived from ExeBench was released~\cite{leach2026obfuscatedexebench}. Whereas our dataset employs OLLVM-based obfuscation, this dataset was generated using Tigress. The two datasets are therefore complementary, and their combined use is recommended.

Rather than proposing a new architecture, we leave \textit{LLM4Decompile} 
unchanged and study obfuscation through the lens of candidate generation 
and selection. We show that multi-candidate sampling with simple 
selectors, combined with fine-tuning on \textit{ExeBench-Obfuscated}, 
already contradicts the unsuitability disclaimer of the original 
\textit{LLM4Decompile} paper. To support follow-up work, we additionally 
release \textit{ExeBench-Obfuscated} as, to the best of our knowledge, the first 
large-scale obfuscated source-to-binary dataset purpose-built for 
training neural decompilers.
    
\bheading{Selection versus Generation}
\label{sec:related-selection}
The observation that the most-probable output of a probabilistic model
is not necessarily the best one dates back to Minimum Bayes-Risk (MBR)
decoding~\cite{kumar-byrne-2004-minimum,eikema-aziz-2020-map}. In LLM reasoning,
selection has been studied through majority voting over sampled chains
of thought~\cite{DBLP:conf/iclr/0002WSLCNCZ23} or process reward
models~\cite{DBLP:conf/iclr/LightmanKBEBLLS24}. 
For code, several concrete selectors have been proposed: 
\emph{AlphaCode} clusters candidates by execution
behaviour~\cite{DBLP:journals/corr/abs-2203-07814},
\emph{MBR-EXEC} chooses the program whose traces agree most with
peers~\cite{DBLP:conf/emnlp/ShiFGZW22} and
\emph{Coder-Reviewer} reranks with a bidirectional likelihood
score~\cite{DBLP:conf/icml/ZhangYHLYF023}. 
A complementary line of research replaces
selection with iterative repair from compiler or runtime
feedback~\cite{DBLP:conf/iclr/ChenLSZ24,DBLP:conf/iclr/OlaussonIW0S24}.
In neural decompilation, by contrast, candidate selection has received little attention. \emph{SLaDe}~\cite{DBLP:conf/cgo/ArmengolEstapeWCO24} ranks samples through executable test cases, a signal similar to our execution guided selector, but unavailable for most binaries in the wild. Most current LLM-based decompilers~\cite{DBLP:conf/emnlp/TanL0Z24/LLM4Decompile,DBLP:journals/corr/abs-2509-14646,refdecompile} only report the output of the greedy decoding process and do not consider the search space of other candidates.
We instead treat candidate selection as the object of study, systematically comparing multiple selectors. To our knowledge, this is the first study of generate-then-select pipelines for neural decompilation. %

\section{Conclusion}
\label{sec:conclusion}

We studied neural decompilation as a two-stage problem of candidate generation and selection. Across our experiments, correct reconstructions are often already present in the model's candidate space; the central challenge is identifying them reliably. Simple selection strategies substantially improve single-output accuracy, and, combined with targeted fine-tuning, make LLM4Decompile effective even on obfuscated binaries.

The broader implication is that progress in neural decompilation need not come solely from larger models or more elaborate generation methods. Our results suggest that the candidate distribution produced by existing models contains considerably more useful information than greedy decoding reveals. Neural decompilation should therefore be treated not only as a generation problem, but also as a search and selection problem. Developing reliable selectors, especially ones that do not depend on execution feedback, may be as important as improving the underlying decompiler.

To support follow-up work, we open-source \textit{ExeBench-Obfuscated}, a corpus of 362,647 paired source-to-binary samples (287,790 training, 74,857 test).
    
\section*{Acknowledgements}
This work has been supported by funding from the \textit{Agentur für Innovation in der Cybersicherheit GmbH} (Cyberagentur, project \href{https://sovereign-project.de/}{\textit{SOVEREIGN}}).
    
\bibliographystyle{plain}
\bibliography{references}
    
\appendix
\section{Selection Methods in Table~\ref{tab:rq2-selection}}
\label{app:rq2-selectors}

For each row of Table~\ref{tab:rq2-selection}, we briefly describe how the selector scores a candidate. Unless noted otherwise, the candidate maximizing the score is returned. Let $x'$ denote the decompiler input (Ghidra pseudo-code for C1/C3, assembly for C2/C4) and $c$ a candidate in  source code. The pool $\mathcal{P}$ is the set of candidates that compile, except where stated.

\begin{description}[leftmargin=0pt, labelindent=0pt, itemindent=0pt]

\item[Oracle.] Returns any $c \in \mathcal{P}$ with $\textsc{passesTests}(c) = \text{true}$, otherwise an arbitrary $c \in \mathcal{P}$. Reported as an upper bound, not deployable.

\item[Neg.\ Perplexity.] Scores $c$ by the negative perplexity of $c$ conditioned on the LLM4Decompile prompt built from $x$, computed token-wise from the model's log-probabilities over the candidate portion only. Lower perplexity is preferred.

\item[MBR Consensus.]Scores each $c \in \mathcal{P}$ and returns the highest expected agreement with other compiling $c$ . Agreement is computed via weighted lexical /structural similarity, and weighted by empirical candidate frequency after light normalization~\cite{kumar-byrne-2004-minimum,eikema-aziz-2020-map}.

\item[Majority Vote (Normalized, Compile Only).] Voting key is $c$ after CRLF normalization and removal of whitespaces. The most frequent key over $\mathcal{P}$ wins.

\item[Majority Vote (Exact, Compile Only).] Same as above but the voting key is the raw candidate string $c$ with no normalization, over $\mathcal{P}$.

\item[Majority Vote (Normalized, All).] Normalized voting key over the full candidate pool (including non-compiling candidates), not $\mathcal{P}$.

\item[Majority Vote (Exact, All).] Exact-string voting key over the full candidate pool.

\item[Execution-Guided.] For each candidate $c \in \mathcal{P}$, every unit test assertion in the sample's test harness is rewritten as a print statement. This turns each pass/fail check into a truth value for each line in the standard output (stdout). Upon execution, the sample is scored based on the number of correct assertions, i.e., assertions that would have evaluated to true.

\item[Random.] Uniformly samples one $c \in \mathcal{P}$.

\item[Ensemble (Text).] Weighted Combination of \emph{Bigram Similarity}, \emph{Trigram Similarity} and \emph{Token F1}.

\item[Bigram Similarity.] Jaccard similarity of token bigrams between $c$ and $x'$, where tokens come from a regex C tokenizer over identifiers, numbers, operators, punctuation, and string/char literals.

\item[Trigram Similarity.] Same as bigram similarity but with token trigrams.

\item[UniXcoder (Fine-tuned).] Scores c by the cosine similarity  between the projected sentence embeddings of $c$ and $x'$ produced by a UniXcoder encoder with a two-layer MLP projection head, fine-tuned contrastively on (input, source) pairs. 

\item[Edit Similarity.] $1 - \mathrm{lev}(c,x)/\max(|c|,|x|)$, where $\mathrm{lev}$ is character-level Levenshtein distance.

\item[Token F1.] F1 score over the set of tokens shared between the regex-tokenized $c$ and $x$ (set semantics, not multiset).

\item[AST Node Similarity.] Tree-sitter is used to parse $c$ and $x'$ as C code. Each node-type label is counted into a vector and the selector returns the cosine similarity of the two count vectors. On C2/C4 the assembly input does not parse as C, so this comparison degrades to the empty tree on one side.

\item[UniXcoder (Base).] Same as the fine-tuned variant but with the pretrained \texttt{microsoft/unixcoder-base} encoder, i.e., without fine-tuning.

\item[Identifier Similarity.] Jaccard similarity of identifier sets extracted from the tree-sitter AST of $c$ and $x$, after removing C keywords and primitive type names. On C2/C4, the assembly input does not parse as C.

\item[Control Flow Similarity.] Counts occurrences of eight tree-sitter node types in $c$ and $x$ (\texttt{if\_statement}, \texttt{for\_statement}, \texttt{while\_statement}, \texttt{do\_statement}, \texttt{return\_statement}, \texttt{call\_expression}, \texttt{binary\_expression}, \texttt{assignment\_\\expression}); score is $1 - \sum_k |n_c^k - n_x^k| / (2 \sum_k \max(n_c^k, n_x^k))$. On C2/C4, the assembly input does not parse as C.

\item[RepE Correctness] Following Ribeiro \textit{et al.}~\cite{ribeiro2026llmsinternalrepresentationcode}, a correctness direction is fit per layer via PCA on hidden-state differences of paired correct/incorrect code stimuli; $c$ is scored by the dot product of its last-token hidden state (at the validation-best layer) with this direction.

\end{description}

\section{Mirai Source Code - Case Study}
\label{app:mirai_case_study}

The \texttt{toggle\_obf} function analyzed in Section~\ref{sec:case-study} is sourced from the \texttt{mirai/bot/\\table.c} source file within the publicly available Mirai repository~\cite{jgamblin_mirai_source_code}. The original code together with its recompiled versions is shown in \Cref{fig:case-study}.

\begin{figure*}[!t]
  \centering

  \begin{subfigure}[t]{0.48\textwidth}
  \centering
  \caption{Ground truth}
  \begin{adjustbox}{max width=\linewidth}
  \begin{minipage}{\linewidth}
\begin{Verbatim}
static void toggle_obf(uint8_t id) {
    int i;
    struct table_value *val = &table[id];
    uint8_t k1 = table_key & 0xff,
            k2 = (table_key >> 8) & 0xff,
            k3 = (table_key >> 16) & 0xff,
            k4 = (table_key >> 24) & 0xff;

    for (i = 0; i < val->val_len; i++)
    {
        val->val[i] ^= k1;
        val->val[i] ^= k2;
        val->val[i] ^= k3;
        val->val[i] ^= k4;
    }

#ifdef DEBUG
    val->locked = !val->locked;
#endif
}
\end{Verbatim}
  \end{minipage}
  \end{adjustbox}
  \end{subfigure}
  \hfill
  \begin{subfigure}[t]{0.48\textwidth}
  \centering
  \caption{Ghidra pseudo-code}
  \begin{adjustbox}{max width=\linewidth}
  \begin{minipage}{\linewidth}
\begin{Verbatim}
void toggle_obf(uint param_1) {
  undefined4 uVar1;
  long lVar2;
  long lVar3;
  byte *pbVar4;
  long *plVar5;
  
  uVar1 = table_key;
  lVar2 = (ulong)(param_1 & 0xff) * 0x10;
  plVar5 = (long *)(table + lVar2);
  if (*(short *)(table + lVar2 + 8) != 0) {
    lVar3 = 0;
    do {
      *(byte *)(*plVar5 + lVar3) = *(byte *)(*plVar5 + lVar3) ^ (byte)uVar1;
# ...
      *pbVar4 = *pbVar4 ^ (byte)((uint)uVar1 >> 0x18);
    } while ((int)lVar3 < (int)(uint)*(ushort *)(table + lVar2 + 8));
  }
  return;
}
\end{Verbatim}
  \end{minipage}
  \end{adjustbox}
  \end{subfigure}

  \vspace{0.8em}

  \begin{subfigure}[t]{0.48\textwidth}
  \centering
  \caption{RQ3 original LLM4Decompile-Ref-9B-v2 model}
  \begin{adjustbox}{max width=\linewidth}
  \begin{minipage}{\linewidth}
\begin{Verbatim}
void toggle_obf(unsigned char ch) {
  int i;
  unsigned int key = table_key;

  if (table[ch].len) {
    for (i = 0; i < table[ch].len; i++) {
      table[ch].obf[i] ^= (key & 0x000000ff);
      table[ch].obf[i] ^= (key & 0x0000ff00) >> 8;
      table[ch].obf[i] ^= (key & 0x00ff0000) >> 16;
      table[ch].obf[i] ^= (key & 0xff000000) >> 24;
    }
  }
}
\end{Verbatim}
  \end{minipage}
  \end{adjustbox}
  \end{subfigure}
  \hfill
  \begin{subfigure}[t]{0.48\textwidth}
  \centering
  \caption{RQ3 fine-tuned perplexity selected}
  \begin{adjustbox}{max width=\linewidth}
  \begin{minipage}{\linewidth}
\begin{Verbatim}
void toggle_obf(int fd) {
    int i;
    unsigned short *p_len;
    unsigned char *p_val;
    unsigned int key = table_key;

    p_len = (unsigned short *)&table[fd][0];
    p_val = (unsigned char *)&table[fd][2];

    for (i = 0; i < *p_len; i++) {
        *p_val ^= key & 0xff;
        *p_val ^= (key >> 8) & 0xff;
        *p_val ^= (key >> 16) & 0xff;
        *p_val ^= (key >> 24) & 0xff;
        p_val++;
    }
}
\end{Verbatim}
  \end{minipage}
  \end{adjustbox}
  \end{subfigure}

 \caption{Recovered versions of \texttt{toggle\_obf} from Ghidra, the off-the-shelf LLM4Decompile-Ref-9B-v2 model, and the fine-tuned RQ3 model with the generate-then-select configuration ($N=100$, perplexity selection).}
  \label{fig:case-study}
  \end{figure*}

\end{document}